\documentclass[aps,prl,reprint,longbibliography]{revtex4-1}

\usepackage{graphicx}
\usepackage{graphics}
\usepackage{amsfonts}
\usepackage{amsmath}
\usepackage{amssymb}
\usepackage{color}
\usepackage{ulem}
\usepackage{ifthen}
\usepackage{bbm}

\definecolor{orange}{rgb}{1,0.5,0}
\definecolor{darkgreen}{rgb}{0,0.4,0.1}

\newcommand{\avg}[1]{\left\langle\displaystyle #1\right\rangle}

\newcommand{\identity}{\mathbbm{1}}
\newcommand{\dzero}{D_\infty}
\newcommand{\dpbc}{\mathbb{D}_{\rm PBC}}

\newcommand{\bfk}{{\bf k}}
\newcommand{\bfc}{{\bf c}}
\newcommand{\bfr}{{\bf r}}
\newcommand{\bfv}{{\bf v}}

\newcommand{\lpara}{L_\parallel}
\newcommand{\lperp}{L_\perp}

\newcommand{\lbx}{\Delta x}
\newcommand{\lbt}{\Delta t}

\newcommand{\bra}{\left[}
\newcommand{\ket}{\right]}

\newcommand{\highlightrevision}{false}

\ifthenelse{\equal{\highlightrevision}{true}}
{

\newcommand{\revbar}[1]{{\sout{#1}}}
}
{

\newcommand{\revbar}[1]{}
}

\begin{document}

\title{Transient hydrodynamic finite size effects 
\\ in simulations under periodic boundary conditions
}

\author{Adelchi J. Asta$^1$}
\author{Maximilien Levesque$^{2,3}$}
\author{Rodolphe Vuilleumier$^{2,3}$}
\author{Benjamin Rotenberg$^{1,4}$}
\affiliation{$^1$ \small Sorbonne Universit\'es, UPMC Univ Paris 06, CNRS, UMR 8234 PHENIX, 
4 Place Jussieu, 75005 Paris, France}
\affiliation{$^2$ \small Ecole Normale Sup\'erieure, PSL Research University,
UPMC Univ Paris 06, CNRS, D\'epartement de Chimie, PASTEUR, 24 rue Lhomond,
75005 Paris, France}
\affiliation{$^3$ \small Sorbonne Universit\'es, UPMC Univ Paris 06, ENS, CNRS, 
PASTEUR, 75005 Paris, France}
\affiliation{$^4$ \small R\'eseau sur le Stockage Electrochimique de l'Energie (RS2E), FR CNRS 3459, France}

\date{\today}

\begin{abstract}
We use Lattice-Boltzmann and analytical calculations to investigate
transient hydrodynamic finite size effects induced by the use of periodic
boundary conditions in simulations at the molecular,
mesoscopic or continuum levels of description.
We analyze the transient response to a local 
perturbation in the fluid and obtain via linear response theory the 
local velocity correlation function. 
This new approach is validated by comparing the finite size
effects on the steady-state velocity with the known results for
the diffusion coefficient.
We next investigate the full time-dependence of
the local velocity auto-correlation function. 
We find at long times a cross-over between the expected $t^{-3/2}$ hydrodynamic tail
and an oscillatory exponential decay, and study the scaling
with the system size of the cross-over time, exponential rate and
amplitude, and oscillation frequency. We interpret these results from
the analytic solution of the compressible Navier-Stokes equation 
for the slowest modes, which are set by the system size.
The present work not only provides a comprehensive analysis
of hydrodynamic finite size effects in bulk fluids, but also
establishes the Lattice-Boltzmann method as a suitable tool to investigate 
such effects in general.
\end{abstract}

\maketitle

It is by now well established that hydrodynamic finite size effects arise
in simulations due to the use of periodic boundary conditions (PBC). These 
effects can be understood as the result of spurious hydrodynamic interactions 
between particles and their periodic images. 
Following D\"unweg and Kremer~\cite{dunweg_molecular_1993},
Yeh and Hummer~\cite{yeh_system-size_2004} proposed a complete
analysis of the finite size effect on the diffusion coefficient 
of fluid particles in a cubic box based on the mobility tensor $\mathbb{T}$:
\begin{align}
\label{eq:dpbc}
\dpbc &=\dzero \identity + k_BT \lim_{r\to 0} 
\left[ \mathbb{T}_{\rm PBC}(\bfr)-\mathbb{T}_\infty(\bfr) \right]
\, ,
\end{align}
with $k_B$ Boltzmann's constant and $T$ the temperature and
where PBC and $\infty$ subscript denote properties 
under periodic and unbounded conditions, respectively,
while $\identity$ is the identity matrix. 
This results in a finite size scaling of the diffusion constant 
$D(L)=D_\infty-\xi k_BT/6\pi\eta L$ for a cubic box of size $L$,
with $\xi\approx2.837$ a constant 
and $\eta$ the fluid viscosity.
The same scaling was found independently~\cite{fushiki_system_2003} and
has been confirmed in molecular dynamics simulations of 
a number of simple fluids~\cite{yeh_system-size_2004}, including several water 
models~\cite{tazi_diffusion_2012,rozmanov_transport_2012},
ionic liquids~\cite{gabl_computational_2012}
or more complex fluids such as solutions of star
polymers~\cite{singh_hydrodynamic_2014}.
More recently, the extension to anisotropic boxes was also 
investigated~\cite{kikugawa_effect_2015,kikugawa_hydrodynamic_2015}
and interpreted in terms of the same hydrodynamic
arguments~\cite{botan_diffusion_2015,vogele_divergent_2016}.

The distortion of the flow field due to the finite size of the system 
(and the associated use of PBC) does not only affect 
the diffusion coefficient $D$ 
of particles, but in principle all dynamical properties.
In particular, hydrodynamic flows in an unbounded fluid
result in long-time tails of correlations functions,
\textit{e.g.} as $t^{-3/2}$ for the velocity autocorrelation function (VACF) in three 
dimensions~\cite{ernst_asymptotic_1971,HansenBook}. 
Such long time tails have been reported 
in molecular simulations for the VACF since the pioneering work of 
Ref.~\citenum{alder_velocity_1967} (see \textit{e.g.}~\cite{levesque_long-time_1974})
as well as in purely hydrodynamic lattice simulations for the VACF or
other correlation functions~\cite{van_der_hoef_long-time_1990,lowe_long-time_1995,
van_der_hoef_computer_1995,lowe_super_1995}.
Such slow hydrodynamic modes also manifest themselves in the 
non-Markovian dynamics of solutes, which includes a deterministic
component of the force exerted by the suspending fluid,
well described for colloidal spheres by the Basset-Boussinesq 
force~\cite{boussinesq_theorie_1901,chow_effect_1972}. Simulations displaying
such a hydrodynamic memory,
either on a coarse-grained~\cite{franosch_resonances_2011} or
molecular~\cite{lesnicki_molecular_2016}
scale, may therefore suffer from artefacts associated with the use
of PBC, at least on long time scales. This was already recognized
by Alder and Wainwright in their seminal paper where they reported their
results ``up to the time where serious interference between neighbouring
periodically repeated systems is indicated''~\cite{alder_decay_1970}.

Here we address this issue of finite size effects on the transient regime 
by revisiting the above hydrodynamic
approach. We investigate the transient response to a singular perturbation of
the fluid, previously considered to predict the steady-state
mobility~\cite{hasimoto_periodic_1959,yeh_system-size_2004}.
More precisely, we determine numerically the time-dependent Green's function
for the Navier-Stokes equation using Lattice Boltzmann
simulations~\cite{SucciBook}. We validate this new approach 
in the steady-state by comparison with known results,
before turning to finite size effects on the transient hydrodynamic response.
We show that the multiple features of these finite size effects can be
rationalized analytically by considering the decay of the relevant hydrodynamic modes.

The dynamics of an incompressible fluid of mass density $\rho_m$
and shear viscosity $\eta$ can be described by the 
mass conservation $\partial_t\rho_m+\rho_m\nabla\cdot\bfv=0$
and Navier-Stokes (NS) equation:
\begin{align}
\label{eq:NS}
\rho_m\frac{\partial\bfv}{\partial t}
+ \rho_m(\bfv\cdot\nabla)\bfv
&= \eta\nabla^2\bfv -\nabla p + \mathbf{f}
\end{align} 
where $\bfv$ is the velocity field, $p$ is the pressure and
$\mathbf{f}$ is a force density.
In the limit of small Reynolds number, defined as
$Re=\frac{||\rho_m(\bfv\cdot\nabla)\bfv||}{||\eta\nabla^2\bfv||}
\sim\frac{uL}{\nu}$ with $u$ and $L$ the typical velocity and length,
and $\nu=\eta/\rho_m$ the kinematic visscosity,
the expression of both tensors in Eq.~(\ref{eq:dpbc})
can be obtained by determining the Green's function 
for the Stokes equation. 
This corresponds to a vanishing
left hand side in Eq.~(\ref{eq:NS}) and a perturbation:
\begin{align}
\label{eq:perturb}
\mathbf{f}(\bfr') &= \left[\delta(\bfr'-\bfr)-\frac{1}{V} \right]
\, \mathbf{F}
\; ,
\end{align} 
with $\delta$ the Dirac distribution,
$\mathbf{F}$ a force and $V$ the volume of the system,
\textit{i.e.} a singular point force at $\bfr$ and a uniform compensating
background, applied on a fluid initially at rest.
The mobility tensor then follows from the steady state velocity 
as $\bfv(\bfr')=\mathbb{T}(\bfr',\bfr)\cdot{\bf F}$.
Note that the limit in Eq.~(\ref{eq:dpbc}) corresponds to $\bfr'\to\bfr$.
The result for the unbounded case is the well-known
Oseen tensor $\mathbb{T}_\infty(\bfr)=\frac{1}{8\pi\eta r}
\left( \identity + \frac{\bfr\bfr}{r^2} \right)$,
while under PBC it is more conveniently expressed in Fourier
space~\cite{yeh_system-size_2004}.

Similarly, the full dynamical response can be obtained by considering
a perturbation of the form $\mathbf{f}(\bfr')\Theta(t)$,
where $\Theta(t)$ is the Heaviside function and the spatial dependence
is given by Eq.~(\ref{eq:perturb}), applied on a fluid initially at rest. 
The Green's function for the time-dependent NS equation,
which corresponds to a perturbation $\mathbf{f}(\bfr')\delta(t)$
is obtained as the time-derivative of the solution $\bfv(r',t)$.
In the limit $Re\ll1$,
the response to $\mathbf{f}(\bfr')\Theta(t)$ converges at long times toward 
the stationary field corresponding to the mobility tensor.

The transient hydrodynamic regime, as quantified by the Green's function,
is also related to the equilibrium fluctations of the velocity field.
Using linear response theory~\cite{HansenBook}, it is easy to show that the 
average velocity $v$ (in the sense of a canonical average over initial
configurations) in the direction of the force
at the position where it is applied, evolves as:
$\frac{d}{dt}\avg{v(\bfr,t)} =
\frac{1}{k_B T} \int d\bfr' \avg{ v(\bfr,t) v(\bfr',0)}f(\bfr',t)$.
This simplifies for the perturbation considered in Eq.~(\ref{eq:perturb}), 
since the total applied force vanishes and so does the total flux
$\int d\bfr' v(\bfr',0)$. One can finally express the velocity
auto-correlation of the local velocity field (LVACF) as:
\begin{align}
\label{eq:lvacf}
Z(t) &\equiv \avg{ v(\bfr,t) v(\bfr,0)} =
\frac{k_B T}{F} \frac{d\avg{v(\bfr,t)}}{dt}  
\;,
\end{align}
where $\avg{v(\bfr,t)}$ is the response to the perturbation
Eq.~(\ref{eq:perturb}).
Note that this expression, although similar to the one for the velocity of
a particle under a constant force $F$, has in fact a very different meaning: 
Here a perturbation is applied at a fixed position $\bfr$ 
(together with the compensating background) 
and the fluid velocity is followed at that position.
Integrating between 0 and infinity,
one obtains the steady state velocity 
$v_\infty(\bfr)=\lim_{t\to\infty}\avg{v(\bfr,t)}
= \frac{F}{k_BT}\int_0^\infty\avg{ v(\bfr,t) v(\bfr,0)} {\rm d}t$.
This relation is analoguous to Einstein's relation for 
the mobility of a particle, $\mu=v/F=D/k_BT$, 
with the diffusion coefficient $D=\int_0^\infty Z(t) {\rm d}t$.
In the following we will therefore refer to the integral of
the LVACF as the diffusion coefficient.

\begin{figure}[ht!]
\begin{center}
\vspace{-0.0cm}
\includegraphics[width=8.5cm]{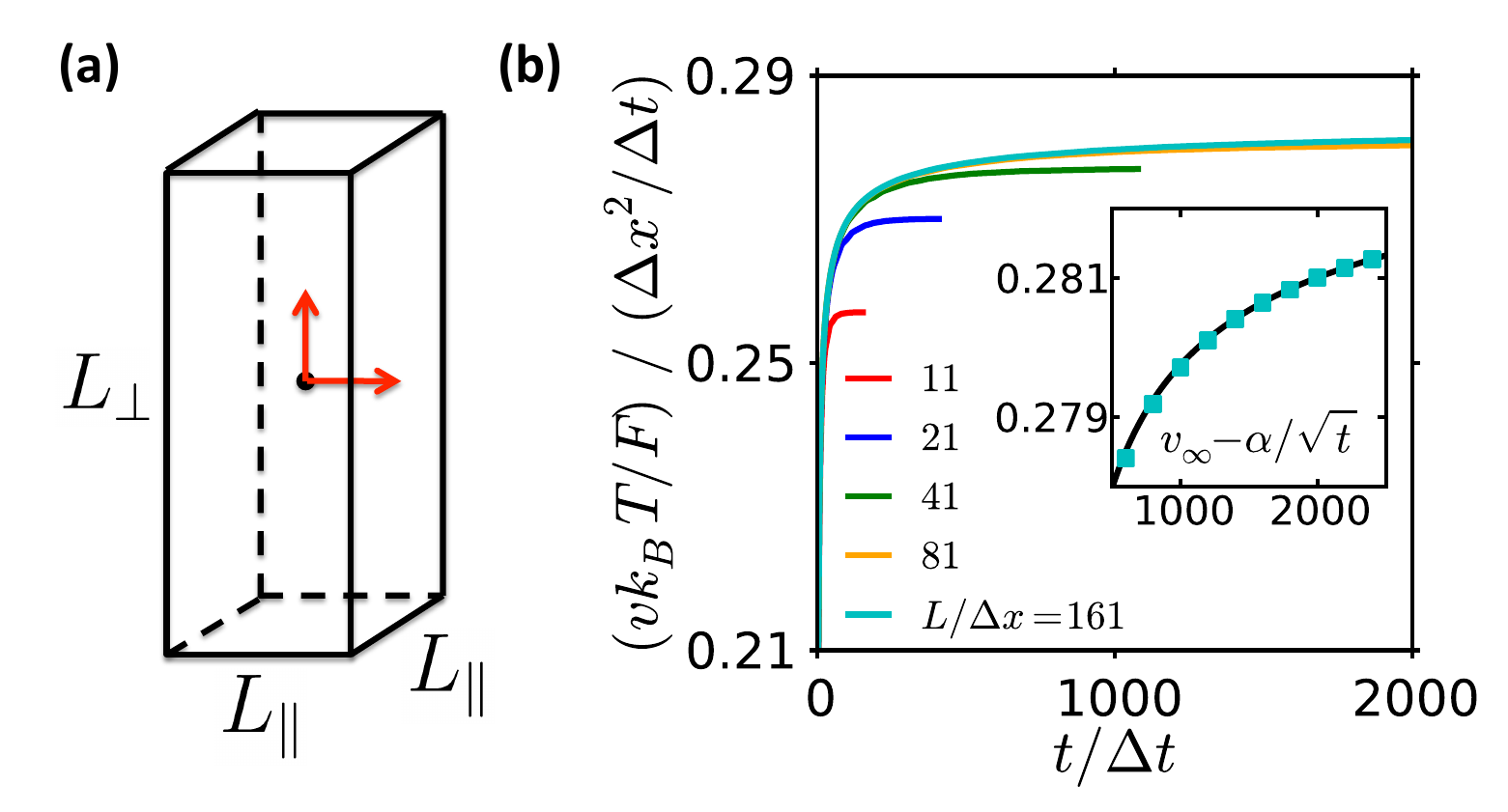}
\vspace{-0.7cm}
\end{center}
\caption{\label{fig:cellvel}
a. A bulk fluid in an orthorhombic cell with one length different
from the other two is submitted to a perturbation Eq.~(\ref{eq:perturb})
which corresponds to a singular point force (in one of the two
relevant directions indicated by red arrows) and a uniform compensating
background. Both elongated ($\lperp>\lpara$, as shown) and flat
($\lperp<\lpara$) boxes are considered.
b. Velocity at point $\bfr$ where the perturbation is applied,
as a function of time, from Lattice Boltzmann simulations
with various cubic boxes of size $L=L_{\perp,\parallel}$.
The inset shows the scaling at long times
used to extrapolate the steady-state velocity,
for the largest system.
}
\end{figure}

Here we use Lattice Boltzmann (LB) simulations~\cite{SucciBook} 
to solve the above hydrodynamic problem, \textit{i.e.} the Navier-Stokes equation
for a fluid initially at rest on which the
perturbation Eq.~(\ref{eq:perturb}) is applied.
In a nutshell, the LB method evolves the
one-particle velocity distribution $f(\bfr,\bfc,t)$
from which the hydrodynamic moments (density, flux, stress tensor)
can be computed. In practice, a kinetic equation is discretized
in space (lattice spacing $\lbx$) and time (time step $\lbt$) and 
so are the velocities, which belong to a finite set $\{\bfc_i\}$
(here we use the D3Q19 lattice). The populations 
$f_i(\bfr,t)\equiv f(\bfr,\bfc_i,t)$ are updated following:
\begin{align}
\label{eq:lb}
f_i(\bfr+\bfc_i\lbt,t+\lbt)&=f_i(\bfr,t)
\nonumber \\
&\hskip-2cm -\frac{\lbt}{\tau}\bra f_i(\bfr,t)-f_i^{eq}(\bfr,t)\ket
+F_i^{ext}(\bfr,t)
\;,
\end{align}
where $f_i^{eq}(\bfr,t)$ corresponds to the local Maxwell-Boltzmann
equilibrium with density $\rho(\bfr,t)=\sum_iw_if_i(\bfr,t)$
and flux $\rho\bfv(\bfr,t)=\sum_iw_if_i(\bfr,t)\bfc_i$,
where $w_i$ are the weights associated with each velocity. 
The relaxation time $\tau$ controls the fluid viscosity.
Here we use $\tau=\lbt$, which results in a kinematic viscosity
$\nu=\frac{c_s^2\lbt}{2}=\frac{1}{6}\frac{\lbx^2}{\lbt}$
since for the D3Q19 lattice the speed of sound is
$c_s=\frac{1}{\sqrt{3}}\frac{\lbx}{\lbt}$
and $F_i^{ext}$ accounts for the external force acting on the 
fluid~\cite{SucciBook}.
We perform simulations for orthorhombic cells with one
length ($\lperp$) different from the other two ($\lpara$),
as illustrated in Figure~\ref{fig:cellvel}a, 
in order to analyze the effect of both the system size and shape.
In practice, we apply the singular perturbation to a single node of
the lattice (and the compensating background everywhere)
and monitor the fluid velocity on that node, as shown
in Figure~\ref{fig:cellvel}b.

\begin{figure}[ht!]
\begin{center}
\vspace{-0.25cm}
\includegraphics[width=8.5cm]{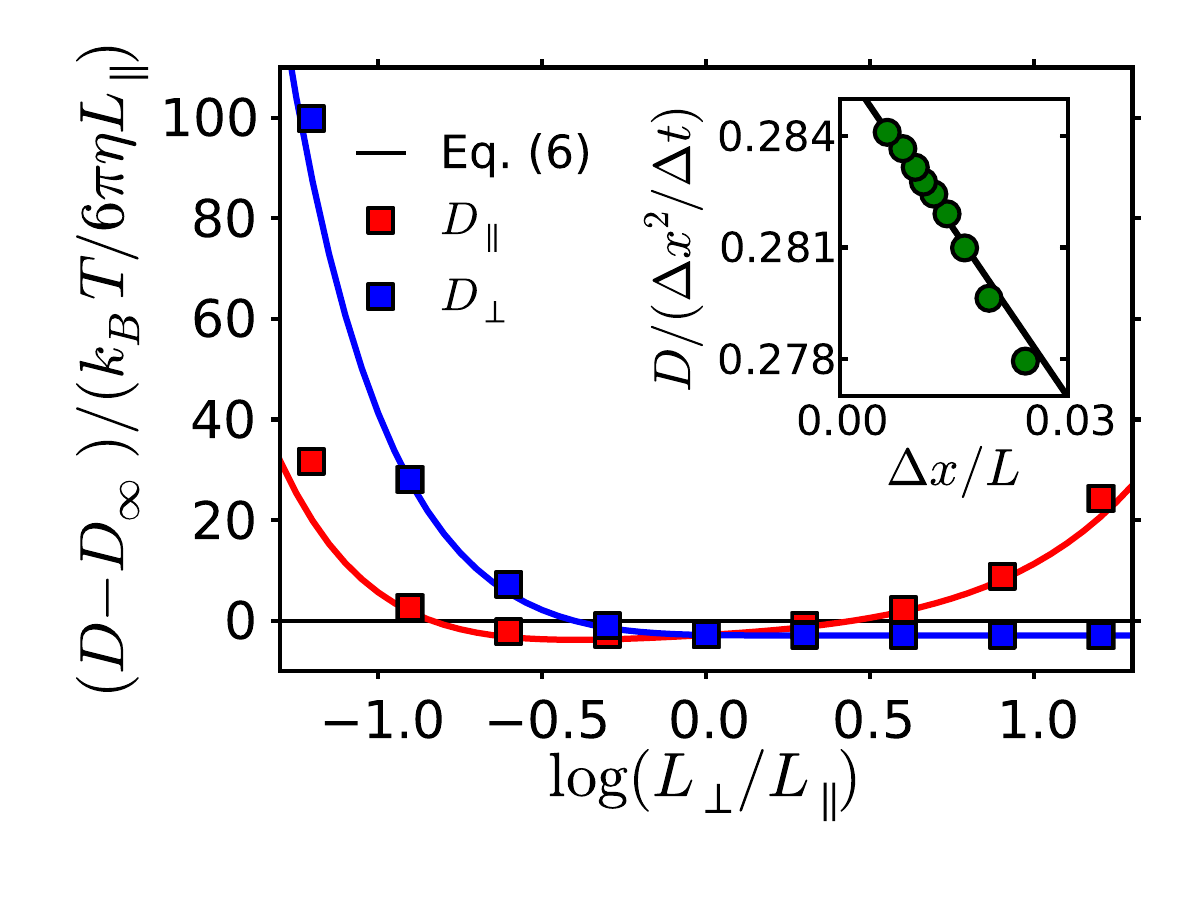}
\vspace{-1.cm}
\end{center}
\caption{\label{fig:diffusion}
Scaling functions
$h_{\parallel,\perp}=(D_{\parallel,\perp}-\dzero)/(k_BT/6\pi\eta\lpara)$
defined in Eq.~(\ref{eq:scalinganiso})
as a function of the aspect ratio $\alpha=\lperp/\lpara$. 
Lattice Bolzmann results (symbols) are compared to 
analytical results (lines) from
Ref.~\citenum{botan_diffusion_2015}.
Note the logarithmic scale on the $x$-axis.
Each point corresponds to the slope of a scaling with system size
at fixed aspect ratio, as illustrated in the inset
for a cubic box ($L=L_{\perp,\parallel}$), 
where the line again corresponds to Eq.~(\ref{eq:scalinganiso}).
}
\end{figure}

As a validation of this new approach for the computation of hydrodynamic
Green's function with LB simulations and linear response, we first
describe the results for the diffusion coefficient, obtained
from the steady-state velocity as $D=k_BTv_\infty(\bfr)/F$.
In the considered geometry, the diffusion tensor is anisotropic
and the two independent components $D_{\parallel,\perp}$
can be determined by applying the perturbation Eq.~(\ref{eq:perturb})
in the corresponding directions. Continuum hydrodynamics predicts
a scaling with system size~\cite{botan_diffusion_2015}:
\begin{align}
\label{eq:scalinganiso}
D_{\parallel,\perp} &=\dzero + \frac{k_BT}{6\pi\eta\lpara} 
h_{\parallel,\perp}\!\left( \frac{\lperp}{\lpara} \right)
\, ,
\end{align}
where the two functions $h_{\parallel,\perp}$ depend only on the aspect
ratio $\lperp/\lpara$. Both functions have been determined 
in Ref.~\cite{botan_diffusion_2015}. 
For the isotropic case $h_{\parallel,\perp}(1)=-\xi\approx-2.837$.
The inset of Figure~\ref{fig:diffusion} shows the diffusion coefficient
for a cubic box as a function of the size $\lperp=\lpara$.
For reasons that will be discussed below, the velocity 
$\avg{v(\bfr,t)}$ converges slowly to its steady-state value,
as $v_\infty-\alpha/\sqrt{t}$ (see the inset of Figure~\ref{fig:cellvel}b).
Therefore we used a fit to this expression at long times 
in order to determine $v_\infty$ for the larger systems.
For small systems the convergence
is faster so that such an extrapolation is not necessary.

The LB results are in excellent agreement with the slope expected
from Eq.~(\ref{eq:scalinganiso}), even though some deviations are
observed for the smaller box sizes ($\sim10\lbx$) as expected.
The extrapolated value for an infinite system is  
$\dzero\approx0.286\lbx^2/\lbt$.
By performing similar size scalings for various aspect 
ratios~\cite{SuppMat}, 
we can compute the scaling functions $h_{\parallel,\perp}$ 
for both components of the diffusion tensor. The results,
shown in the main part of Figure~\ref{fig:diffusion},
are also in excellent agreement with Eq.~(\ref{eq:scalinganiso}). 
This validates the present approach combining linear response
and LB simulations for the determination of hydrodynamic 
finite size effects on the steady-state dynamics.

We now turn to the finite size effects in the transient regime.
As explained above, the Green's function for the time-dependent
NS equation is obtained from the derivative of the response 
$v(\bfr,t)$ to the perturbation ${\bf f}(\bfr')\Theta(t)$.
More precisely, we discuss here the LVACF defined by Eq.~(\ref{eq:lvacf})
which is proportional to this Green's function and quantifies
the equilibrium hydrodynamic fluctuations.
In an unbounded medium, 
such fluctuations are known to result in the long-time tail of
the VACF of particles in a simple fluid according to~\cite{HansenBook}:
\begin{align}
\label{eq:vacfhydro}
Z_\infty(t) &= \frac{2}{3}\frac{k_BT}{\rho_m}
\left[ 4\pi \nu t \right]^{-3/2} \; .
\end{align}
Mode-coupling theory predicts in fact a scaling with 
$D+\nu$ instead of $\nu$, but here the Green's function is not
associated with the diffusion of a tagged particle, so that the corresponding
diffusion coefficient is not involved.

\begin{figure}[ht!]
\begin{center}
\vspace{-0.2cm}
\includegraphics[width=8.5cm]{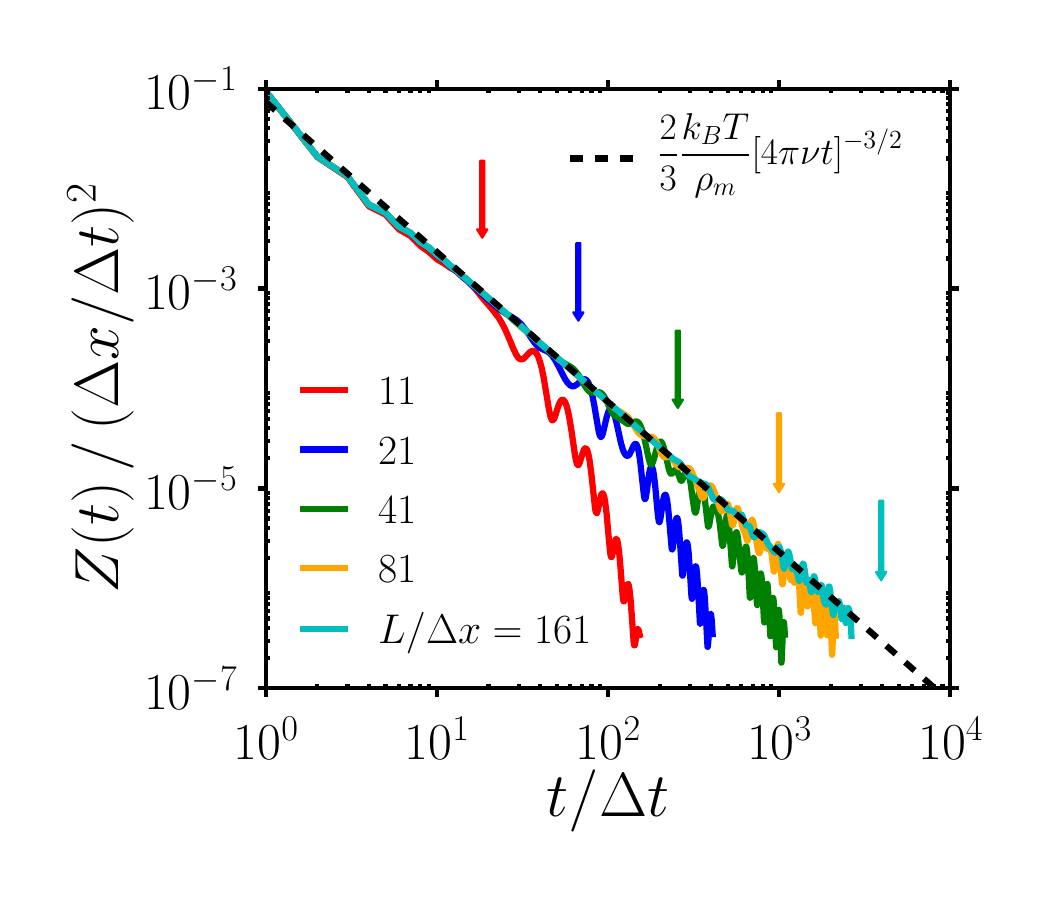}
\vspace{-1cm}
\end{center}
\caption{\label{fig:vacf}
Local velocity auto-correlation function (LVACF) 
computed from Lattice Boltzmann simulations in a cubic cell,
for various cell sizes $L/\lbx$.
The double logarithmic scale underlines the algebraic decay
expected from hydrodynamics in an unbounded fluid, Eq.~(\ref{eq:vacfhydro}).
The finite size results in a cross-over to an exponential
decay, analyzed in further detail in Figure~\ref{fig:vacf2}.
The arrows indicate the diffusion time for the slowest mode,
$\tau_L=1/\nu k_L^2$, with $k_L=2\pi/L$, which corresponds to the
exponential decay rate and is also typical of the cross-over between the
algebraic and exponential regimes.
}
\end{figure}

Figure~\ref{fig:vacf} reports the LVACF computed from Eq.~(\ref{eq:lvacf}) 
using the present LB approach, for various cubic boxes of size $L$.
For the larger systems, the simulation results coincide exactly with
the hydrodynamic scaling Eq.~(\ref{eq:vacfhydro}) over several 
orders of magnitude, without any ajustable parameter.
This scaling, together with Eq.~(\ref{eq:lvacf}), justifies a posteriori
the fit of the velocity as $v_\infty-\alpha/\sqrt{t}$ to extrapolate
the steady-state value.
However, we observe a cross-over from the algebraic decay
to an exponential regime (and oscillations discussed in more detail below), 
with a cross-over time that decreases with decreasing $L$.

It is useful to recall that the algebraic decay Eq.~(\ref{eq:vacfhydro})
results from the superposition of an infinite number of modes (corresponding
to the hydrodynamic limit of vanishing wave numbers $k\to0$) 
for momentum diffusion, which in Fourier space decay
as $\sim e^{-\nu k^2t}$. 
The exponential decay therefore results from the cut-off at low wave 
numbers introduced by the PBC, with the slowest mode corresponding to
$k_L=2\pi/L$ and a characteristic time $\tau_L=1/\nu k_L^2$.
The vertical arrows in Figure~\ref{fig:vacf}, which indicate this
time, show that it is also typical of the cross-over from algebraic to
exponential decay $A_L e^{-\nu k_L^2t}$ of the LVACF.
The prefactor $A_L$ can be roughly estimated
by assuming the continuity between the two regimes at $t=\tau_L$. 
Using Eq.~(\ref{eq:vacfhydro}), this results in: 
\begin{align}
\label{eq:crossoveramplitude}
A_L &= \frac{2e}{3\left[ 4\pi \right]^{3/2}}\frac{k_BT}{\rho_m}
k_L^3
\; .
\end{align}

Another striking feature of the results in Figure~\ref{fig:vacf} 
is the presence of oscillations, with a frequency which depends
on the size of the simulation box. This is clearly another finite size
effect, which can be understood in terms of the slight compressibility
of the fluid. Indeed, in the LB method the fluid is only
quasi-incompressible. 
In such a case, while the transverse mode decays as $\sim e^{-\nu k^2t}$
(as for an incompressible fluid), 
the longitudinal modes follow a dispersion relation which 
can be obtained by linearizing the mass conservation and compressible
NS equation, for an isothermal perturbation of the form 
$e^{i(\omega t - \bfk\cdot\bfr)}$. Using the fact that the equation of state
of the LB fluid is that of an ideal gas ($p=\rho k_BT=\rho_mc_s^2$), one obtains
the following dispersion relation:
$(i\omega)^2+i\omega k^2\left(\frac{4}{3}\nu+\nu'\right)+c_s^2k^2=0$,
with $\nu'=\zeta/\rho_m$ the kinematic bulk viscosity.
In the case of the D3Q19 lattice, for which $\nu'=\frac{2}{3}\nu$,
the solutions are of the form
$i\omega=-\nu k^2\pm ikc_s\sqrt{1-\nu^2k^2/c_s^2}
\sim-\nu k^2\pm ikc_s$ (for $k\ll c_s/\nu$),
\textit{i.e.}\ attenuated sound waves. Such a dispersion relation
had already been considered for the LB simulation of acoustic
waves, see \textit{e.g.}~\cite{dellar_bulk_2001,li_lattice_2011,viggen_viscously_2011}.

\begin{figure}[ht!]
\begin{center}
\vspace{-0.cm}
\includegraphics[width=8.cm]{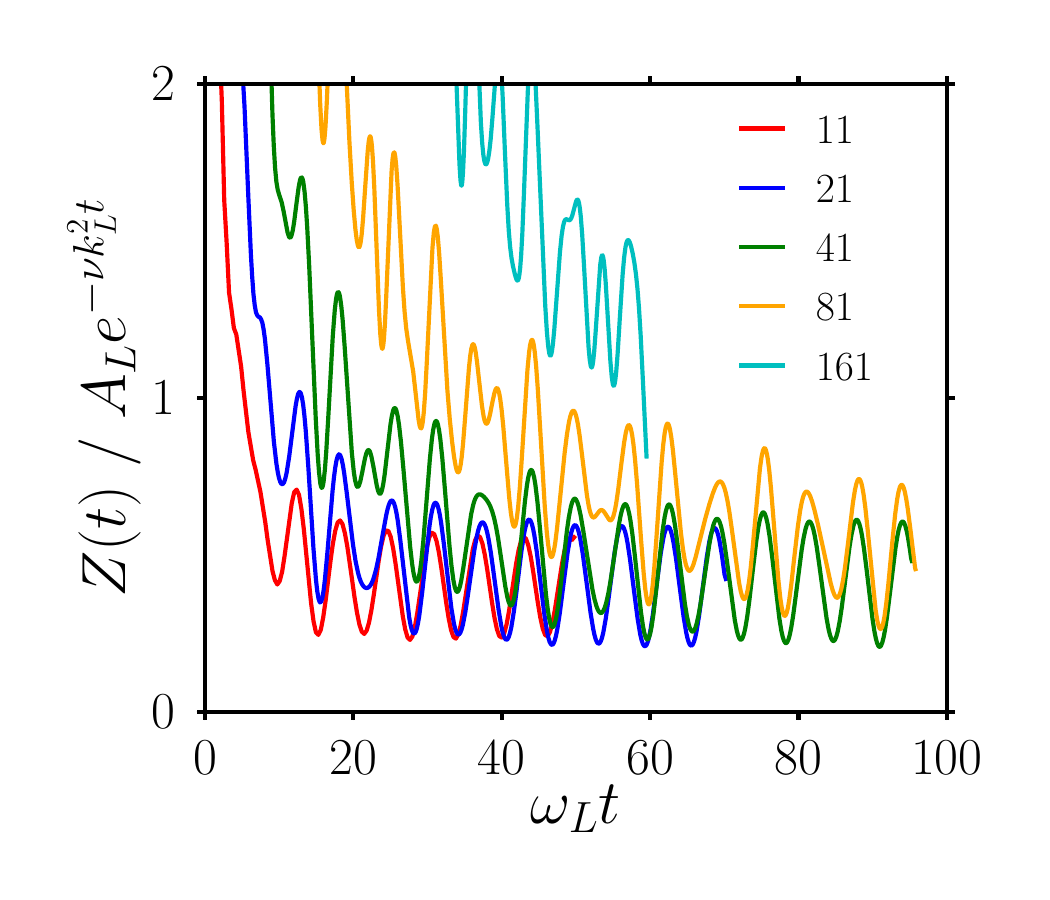}
\vspace{-1cm}
\end{center}
\caption{\label{fig:vacf2}
Local velocity autocorrelation function 
computed from Lattice Boltzmann simulations in a cubic cell,
for various cell sizes $L/\lbx$,
normalized by the expected exponential scaling at long times
$A_L e^{-\nu k_L^2t}$ with $k_L=2\pi/L$ and
$A_L$ given by Eq.~(\ref{eq:crossoveramplitude}), as a function of
time rescaled by the frequency $\omega_L=kc_s\sqrt{1-\nu^2k^2/c_s^2}$.
The oscillations are due to the small compressibility of the Lattice-Boltzmann
fluid, which results in damped acoustic waves.
For the larger systems the contribution of slower modes
$nk_L$ is still visible on the time scale of the simulations.
}
\end{figure}

For periodic systems, the slowest modes correspond to
$k_L=2\pi/L$ and longitudinal modes decay as $\sim e^{-\nu k_L^2t}\cos{\omega_Lt}$, 
with a frequency $\omega_L=k_Lc_s\sqrt{1-\nu^2k_L^2/c_s^2}\sim k_Lc_s$.
Figure~\ref{fig:vacf2} reports the LVACF normalized by the exponential 
decay $A_L e^{-\nu k_L^2t}$, as a function of the rescaled time $\omega_Lt$,
for various system sizes spanning more than one order of magnitude.
The numerical results clearly show that the above analysis captures
all the main features of the finite size effects on the transient hydrodynamic
response:  1) the rate of the exponential decay, 
since at long times the curves oscillate around a plateau; 
2) the order of magnitude $A_L$ of the exponential regime, since the value
of the plateau is the same for all system sizes; 3) the frequency of the 
oscillations, which are in phase after rescaling by $\omega_L$.
While only the slowest mode contributes to the oscillations for the smallest
system ($L=11\lbx$), others are increasingly visible in this time range
as the system size increases. Indeed, the other modes $nk_L$ decay
as $\sim e^{-\nu n^2 k_L^2t}= e^{-\frac{\nu n^2 k_L^2}{\omega_L} \omega_L t}$,
\textit{i.e.}~$\frac{\nu n^2 k_L^2}{\omega_L} = n^2\frac{\nu k_L}{c_s}$
times faster -- a difference which decreases with increasing $L$.

Overall, the present work shows that it is possible to rationalize all
finite size effects in terms of the cut-off of hydrodynamic modes at small wave 
numbers introduced by the use of PBC.
Coming back to Alder and Wainwright's quote~\cite{alder_decay_1970}, 
the time where neighbouring periodically 
repeated systems seriously interfere corresponds to
momentum diffusion time for the slowest mode, $\tau_L=1/\nu k_L^2$.
It is crucial for the setup and analysis of molecular simulations to
have a good control of these finite size effects, which can be efficiently
computed from the present approach combining linear response and
Lattice-Boltzmann simulations.
In turn, such an analysis is useful to extrapolate the macroscopic 
limit without actually performing the simulations for too large systems.
One could exploit these effects further to determine material properties, 
not only the viscosity from the slope of the diffusion coefficient vs inverse
box size (as for water in first principles molecular dynamics
simulations~\cite{kuhne_static_2009}), but also \textit{e.g.} 
the speed of sound from the oscillation frequency of the LVACF, as shown here.

The systematic finite size analysis of the transient response of bulk fluids 
could also be extended to other situations, such as fluids under confinement 
or near interfaces. For example, the long-time decay of the VACF
under confinement or near a boundary, in an otherwise unbounded fluid, 
scales as $t^{-5/2}$ instead of $t^{-3/2}$ in the
bulk~\cite{hagen_algebraic_1997,huang_effect_2015},
but PBC in the directions parallel to the interface
will also result in deviations from the algebraic decay, as demonstrated here. 
Similarly, it has been recently shown that 
the diffusion coefficient of lipids and carbon nanotubes embedded 
in a membrane diverges logarithmically with system size~\cite{vogele_divergent_2016}
and one should also observe the impact of PBC on the transient dynamics.
This may also prove important for extracting from finite size simulations
other dynamical properties for which hydrodynamics play an important role,
such as memory kernels~\cite{lesnicki_molecular_2016}.



\begin{thebibliography}{33}%
\makeatletter
\providecommand \@ifxundefined [1]{%
 \@ifx{#1\undefined}
}%
\providecommand \@ifnum [1]{%
 \ifnum #1\expandafter \@firstoftwo
 \else \expandafter \@secondoftwo
 \fi
}%
\providecommand \@ifx [1]{%
 \ifx #1\expandafter \@firstoftwo
 \else \expandafter \@secondoftwo
 \fi
}%
\providecommand \natexlab [1]{#1}%
\providecommand \enquote  [1]{``#1''}%
\providecommand \bibnamefont  [1]{#1}%
\providecommand \bibfnamefont [1]{#1}%
\providecommand \citenamefont [1]{#1}%
\providecommand \href@noop [0]{\@secondoftwo}%
\providecommand \href [0]{\begingroup \@sanitize@url \@href}%
\providecommand \@href[1]{\@@startlink{#1}\@@href}%
\providecommand \@@href[1]{\endgroup#1\@@endlink}%
\providecommand \@sanitize@url [0]{\catcode `\\12\catcode `\$12\catcode
  `\&12\catcode `\#12\catcode `\^12\catcode `\_12\catcode `\%12\relax}%
\providecommand \@@startlink[1]{}%
\providecommand \@@endlink[0]{}%
\providecommand \url  [0]{\begingroup\@sanitize@url \@url }%
\providecommand \@url [1]{\endgroup\@href {#1}{\urlprefix }}%
\providecommand \urlprefix  [0]{URL }%
\providecommand \Eprint [0]{\href }%
\providecommand \doibase [0]{http://dx.doi.org/}%
\providecommand \selectlanguage [0]{\@gobble}%
\providecommand \bibinfo  [0]{\@secondoftwo}%
\providecommand \bibfield  [0]{\@secondoftwo}%
\providecommand \translation [1]{[#1]}%
\providecommand \BibitemOpen [0]{}%
\providecommand \bibitemStop [0]{}%
\providecommand \bibitemNoStop [0]{.\EOS\space}%
\providecommand \EOS [0]{\spacefactor3000\relax}%
\providecommand \BibitemShut  [1]{\csname bibitem#1\endcsname}%
\let\auto@bib@innerbib\@empty
\bibitem [{\citenamefont {D\"unweg}\ and\ \citenamefont
  {Kremer}(1993)}]{dunweg_molecular_1993}%
  \BibitemOpen
  \bibfield  {author} {\bibinfo {author} {\bibfnamefont {Burkhard}\
  \bibnamefont {D\"unweg}}\ and\ \bibinfo {author} {\bibfnamefont {Kurt}\
  \bibnamefont {Kremer}},\ }\bibfield  {title} {\enquote {\bibinfo {title}
  {Molecular dynamics simulation of a polymer chain in solution},}\ }\href@noop
  {} {\bibfield  {journal} {\bibinfo  {journal} {The Journal of chemical
  physics}\ }\textbf {\bibinfo {volume} {99}},\ \bibinfo {pages} {6983--6997}
  (\bibinfo {year} {1993})}\BibitemShut {NoStop}%
\bibitem [{\citenamefont {Yeh}\ and\ \citenamefont
  {Hummer}(2004)}]{yeh_system-size_2004}%
  \BibitemOpen
  \bibfield  {author} {\bibinfo {author} {\bibfnamefont {In-Chul}\ \bibnamefont
  {Yeh}}\ and\ \bibinfo {author} {\bibfnamefont {Gerhard}\ \bibnamefont
  {Hummer}},\ }\bibfield  {title} {\enquote {\bibinfo {title} {System-{Size}
  {Dependence} of {Diffusion} {Coefficients} and {Viscosities} from {Molecular}
  {Dynamics} {Simulations} with {Periodic} {Boundary} {Conditions}},}\
  }\href@noop {} {\bibfield  {journal} {\bibinfo  {journal} {The Journal of
  Physical Chemistry B}\ }\textbf {\bibinfo {volume} {108}},\ \bibinfo {pages}
  {15873--15879} (\bibinfo {year} {2004})}\BibitemShut {NoStop}%
\bibitem [{\citenamefont {Fushiki}(2003)}]{fushiki_system_2003}%
  \BibitemOpen
  \bibfield  {author} {\bibinfo {author} {\bibfnamefont {M.}~\bibnamefont
  {Fushiki}},\ }\bibfield  {title} {\enquote {\bibinfo {title} {System size
  dependence of the diffusion coefficient in a simple liquid},}\ }\href@noop {}
  {\bibfield  {journal} {\bibinfo  {journal} {Physical Review E}\ }\textbf
  {\bibinfo {volume} {68}} (\bibinfo {year} {2003})}\BibitemShut {NoStop}%
\bibitem [{\citenamefont {Tazi}\ \emph {et~al.}(2012)\citenamefont {Tazi},
  \citenamefont {Bo{\c t}an}, \citenamefont {Salanne}, \citenamefont {Marry},
  \citenamefont {Turq},\ and\ \citenamefont {Rotenberg}}]{tazi_diffusion_2012}%
  \BibitemOpen
  \bibfield  {author} {\bibinfo {author} {\bibfnamefont {S.}~\bibnamefont
  {Tazi}}, \bibinfo {author} {\bibfnamefont {A.}~\bibnamefont {Bo{\c t}an}},
  \bibinfo {author} {\bibfnamefont {M.}~\bibnamefont {Salanne}}, \bibinfo
  {author} {\bibfnamefont {V.}~\bibnamefont {Marry}}, \bibinfo {author}
  {\bibfnamefont {P.}~\bibnamefont {Turq}}, \ and\ \bibinfo {author}
  {\bibfnamefont {B.}~\bibnamefont {Rotenberg}},\ }\bibfield  {title} {\enquote
  {\bibinfo {title} {Diffusion coefficient and shear viscosity of rigid water
  models},}\ }\href@noop {} {\bibfield  {journal} {\bibinfo  {journal} {Journal
  of Physics: Condensed Matter}\ }\textbf {\bibinfo {volume} {24}},\ \bibinfo
  {pages} {284117} (\bibinfo {year} {2012})}\BibitemShut {NoStop}%
\bibitem [{\citenamefont {Rozmanov}\ and\ \citenamefont
  {Kusalik}(2012)}]{rozmanov_transport_2012}%
  \BibitemOpen
  \bibfield  {author} {\bibinfo {author} {\bibfnamefont {Dmitri}\ \bibnamefont
  {Rozmanov}}\ and\ \bibinfo {author} {\bibfnamefont {Peter~G.}\ \bibnamefont
  {Kusalik}},\ }\bibfield  {title} {\enquote {\bibinfo {title} {Transport
  coefficients of the {TIP}4p-2005 water model},}\ }\href@noop {} {\bibfield
  {journal} {\bibinfo  {journal} {The Journal of Chemical Physics}\ }\textbf
  {\bibinfo {volume} {136}},\ \bibinfo {pages} {044507} (\bibinfo {year}
  {2012})}\BibitemShut {NoStop}%
\bibitem [{\citenamefont {Gabl}\ \emph {et~al.}(2012)\citenamefont {Gabl},
  \citenamefont {Schröder},\ and\ \citenamefont
  {Steinhauser}}]{gabl_computational_2012}%
  \BibitemOpen
  \bibfield  {author} {\bibinfo {author} {\bibfnamefont {Sonja}\ \bibnamefont
  {Gabl}}, \bibinfo {author} {\bibfnamefont {Christian}\ \bibnamefont
  {Schröder}}, \ and\ \bibinfo {author} {\bibfnamefont {Othmar}\ \bibnamefont
  {Steinhauser}},\ }\bibfield  {title} {\enquote {\bibinfo {title}
  {Computational studies of ionic liquids: {Size} does matter and time too},}\
  }\href@noop {} {\bibfield  {journal} {\bibinfo  {journal} {The Journal of
  Chemical Physics}\ }\textbf {\bibinfo {volume} {137}},\ \bibinfo {pages}
  {094501} (\bibinfo {year} {2012})}\BibitemShut {NoStop}%
\bibitem [{\citenamefont {Singh}\ \emph {et~al.}(2014)\citenamefont {Singh},
  \citenamefont {Huang}, \citenamefont {Westphal}, \citenamefont {Gompper},\
  and\ \citenamefont {Winkler}}]{singh_hydrodynamic_2014}%
  \BibitemOpen
  \bibfield  {author} {\bibinfo {author} {\bibfnamefont {Sunil~P.}\
  \bibnamefont {Singh}}, \bibinfo {author} {\bibfnamefont {Chien-Cheng}\
  \bibnamefont {Huang}}, \bibinfo {author} {\bibfnamefont {Elmar}\ \bibnamefont
  {Westphal}}, \bibinfo {author} {\bibfnamefont {Gerhard}\ \bibnamefont
  {Gompper}}, \ and\ \bibinfo {author} {\bibfnamefont {Roland~G.}\ \bibnamefont
  {Winkler}},\ }\bibfield  {title} {\enquote {\bibinfo {title} {Hydrodynamic
  correlations and diffusion coefficient of star polymers in solution},}\
  }\href@noop {} {\bibfield  {journal} {\bibinfo  {journal} {The Journal of
  Chemical Physics}\ }\textbf {\bibinfo {volume} {141}},\ \bibinfo {pages}
  {084901} (\bibinfo {year} {2014})}\BibitemShut {NoStop}%
\bibitem [{\citenamefont {Kikugawa}\ \emph
  {et~al.}(2015{\natexlab{a}})\citenamefont {Kikugawa}, \citenamefont {Ando},
  \citenamefont {Suzuki}, \citenamefont {Naruke}, \citenamefont {Nakano},\ and\
  \citenamefont {Ohara}}]{kikugawa_effect_2015}%
  \BibitemOpen
  \bibfield  {author} {\bibinfo {author} {\bibfnamefont {Gota}\ \bibnamefont
  {Kikugawa}}, \bibinfo {author} {\bibfnamefont {Shotaro}\ \bibnamefont
  {Ando}}, \bibinfo {author} {\bibfnamefont {Jo}~\bibnamefont {Suzuki}},
  \bibinfo {author} {\bibfnamefont {Yoichi}\ \bibnamefont {Naruke}}, \bibinfo
  {author} {\bibfnamefont {Takeo}\ \bibnamefont {Nakano}}, \ and\ \bibinfo
  {author} {\bibfnamefont {Taku}\ \bibnamefont {Ohara}},\ }\bibfield  {title}
  {\enquote {\bibinfo {title} {Effect of the computational domain size and
  shape on the self-diffusion coefficient in a {Lennard}-{Jones} liquid},}\
  }\href@noop {} {\bibfield  {journal} {\bibinfo  {journal} {The Journal of
  chemical physics}\ }\textbf {\bibinfo {volume} {142}},\ \bibinfo {pages}
  {024503} (\bibinfo {year} {2015}{\natexlab{a}})}\BibitemShut {NoStop}%
\bibitem [{\citenamefont {Kikugawa}\ \emph
  {et~al.}(2015{\natexlab{b}})\citenamefont {Kikugawa}, \citenamefont
  {Nakano},\ and\ \citenamefont {Ohara}}]{kikugawa_hydrodynamic_2015}%
  \BibitemOpen
  \bibfield  {author} {\bibinfo {author} {\bibfnamefont {Gota}\ \bibnamefont
  {Kikugawa}}, \bibinfo {author} {\bibfnamefont {Takeo}\ \bibnamefont
  {Nakano}}, \ and\ \bibinfo {author} {\bibfnamefont {Taku}\ \bibnamefont
  {Ohara}},\ }\bibfield  {title} {\enquote {\bibinfo {title} {Hydrodynamic
  consideration of the finite size effect on the self-diffusion coefficient in
  a periodic rectangular parallelepiped system},}\ }\href@noop {} {\bibfield
  {journal} {\bibinfo  {journal} {The Journal of Chemical Physics}\ }\textbf
  {\bibinfo {volume} {143}},\ \bibinfo {pages} {024507} (\bibinfo {year}
  {2015}{\natexlab{b}})}\BibitemShut {NoStop}%
\bibitem [{\citenamefont {Botan}\ \emph {et~al.}(2015)\citenamefont {Botan},
  \citenamefont {Marry},\ and\ \citenamefont
  {Rotenberg}}]{botan_diffusion_2015}%
  \BibitemOpen
  \bibfield  {author} {\bibinfo {author} {\bibfnamefont {Alexandru}\
  \bibnamefont {Botan}}, \bibinfo {author} {\bibfnamefont {Virginie}\
  \bibnamefont {Marry}}, \ and\ \bibinfo {author} {\bibfnamefont {Benjamin}\
  \bibnamefont {Rotenberg}},\ }\bibfield  {title} {\enquote {\bibinfo {title}
  {Diffusion in bulk liquids: finite-size effects in anisotropic systems},}\
  }\href@noop {} {\bibfield  {journal} {\bibinfo  {journal} {Molecular
  Physics}\ }\textbf {\bibinfo {volume} {113}},\ \bibinfo {pages} {2674--2679}
  (\bibinfo {year} {2015})}\BibitemShut {NoStop}%
\bibitem [{\citenamefont {V\"ogele}\ and\ \citenamefont
  {Hummer}(2016)}]{vogele_divergent_2016}%
  \BibitemOpen
  \bibfield  {author} {\bibinfo {author} {\bibfnamefont {Martin}\ \bibnamefont
  {V\"ogele}}\ and\ \bibinfo {author} {\bibfnamefont {Gerhard}\ \bibnamefont
  {Hummer}},\ }\bibfield  {title} {\enquote {\bibinfo {title} {Divergent
  {Diffusion} {Coefficients} in {Simulations} of {Fluids} and {Lipid}
  {Membranes}},}\ }\href@noop {} {\bibfield  {journal} {\bibinfo  {journal}
  {The Journal of Physical Chemistry B}\ }\textbf {\bibinfo {volume} {120}},\
  \bibinfo {pages} {8722Ð8732} (\bibinfo {year} {2016})}\BibitemShut {NoStop}%
\bibitem [{\citenamefont {Ernst}\ \emph {et~al.}(1971)\citenamefont {Ernst},
  \citenamefont {Hauge},\ and\ \citenamefont
  {Van~Leeuwen}}]{ernst_asymptotic_1971}%
  \BibitemOpen
  \bibfield  {author} {\bibinfo {author} {\bibfnamefont {M.~H.}\ \bibnamefont
  {Ernst}}, \bibinfo {author} {\bibfnamefont {E.~H.}\ \bibnamefont {Hauge}}, \
  and\ \bibinfo {author} {\bibfnamefont {J.~M.~J.}\ \bibnamefont
  {Van~Leeuwen}},\ }\bibfield  {title} {\enquote {\bibinfo {title} {Asymptotic
  time behavior of correlation functions. {I}. {Kinetic} terms},}\ }\href@noop
  {} {\bibfield  {journal} {\bibinfo  {journal} {Physical Review A}\ }\textbf
  {\bibinfo {volume} {4}},\ \bibinfo {pages} {2055} (\bibinfo {year}
  {1971})}\BibitemShut {NoStop}%
\bibitem [{\citenamefont {Hansen}\ and\ \citenamefont
  {McDonald}(2013)}]{HansenBook}%
  \BibitemOpen
  \bibfield  {author} {\bibinfo {author} {\bibfnamefont {J.-P.}\ \bibnamefont
  {Hansen}}\ and\ \bibinfo {author} {\bibfnamefont {I.R.}\ \bibnamefont
  {McDonald}},\ }\href@noop {} {\emph {\bibinfo {title} {Theory of Simple
  Liquids, 4th Edition}}}\ (\bibinfo  {publisher} {Academic Press},\ \bibinfo
  {year} {2013})\BibitemShut {NoStop}%
\bibitem [{\citenamefont {Alder}\ and\ \citenamefont
  {Wainwright}(1967)}]{alder_velocity_1967}%
  \BibitemOpen
  \bibfield  {author} {\bibinfo {author} {\bibfnamefont {B.~J.}\ \bibnamefont
  {Alder}}\ and\ \bibinfo {author} {\bibfnamefont {T.~E.}\ \bibnamefont
  {Wainwright}},\ }\bibfield  {title} {\enquote {\bibinfo {title} {Velocity
  autocorrelations for hard spheres},}\ }\href@noop {} {\bibfield  {journal}
  {\bibinfo  {journal} {Physical review letters}\ }\textbf {\bibinfo {volume}
  {18}},\ \bibinfo {pages} {988} (\bibinfo {year} {1967})}\BibitemShut
  {NoStop}%
\bibitem [{\citenamefont {Levesque}\ and\ \citenamefont
  {Ashurst}(1974)}]{levesque_long-time_1974}%
  \BibitemOpen
  \bibfield  {author} {\bibinfo {author} {\bibfnamefont {D.}~\bibnamefont
  {Levesque}}\ and\ \bibinfo {author} {\bibfnamefont {W.T.}\ \bibnamefont
  {Ashurst}},\ }\bibfield  {title} {\enquote {\bibinfo {title} {Long-{Time}
  {Behavior} of the {Velocity} {Autocorrelation} {Function} for a {Fluid} of
  {Soft} {Repulsive} {Particles}},}\ }\href@noop {} {\bibfield  {journal}
  {\bibinfo  {journal} {Physical Review Letters}\ }\textbf {\bibinfo {volume}
  {33}},\ \bibinfo {pages} {277} (\bibinfo {year} {1974})}\BibitemShut
  {NoStop}%
\bibitem [{\citenamefont {van~der Hoef}\ and\ \citenamefont
  {Frenkel}(1990)}]{van_der_hoef_long-time_1990}%
  \BibitemOpen
  \bibfield  {author} {\bibinfo {author} {\bibfnamefont {M.~A.}\ \bibnamefont
  {van~der Hoef}}\ and\ \bibinfo {author} {\bibfnamefont {D.}~\bibnamefont
  {Frenkel}},\ }\bibfield  {title} {\enquote {\bibinfo {title} {Long-time tails
  of the velocity autocorrelation function in two- and three-dimensional
  lattice-gas cellular automata: {A} test of mode-coupling theory},}\
  }\href@noop {} {\bibfield  {journal} {\bibinfo  {journal} {Physical Review
  A}\ }\textbf {\bibinfo {volume} {41}},\ \bibinfo {pages} {4277--4284}
  (\bibinfo {year} {1990})}\BibitemShut {NoStop}%
\bibitem [{\citenamefont {Lowe}\ \emph {et~al.}(1995)\citenamefont {Lowe},
  \citenamefont {Frenkel},\ and\ \citenamefont
  {Masters}}]{lowe_long-time_1995}%
  \BibitemOpen
  \bibfield  {author} {\bibinfo {author} {\bibfnamefont {C.~P.}\ \bibnamefont
  {Lowe}}, \bibinfo {author} {\bibfnamefont {D.}~\bibnamefont {Frenkel}}, \
  and\ \bibinfo {author} {\bibfnamefont {A.~J.}\ \bibnamefont {Masters}},\
  }\bibfield  {title} {\enquote {\bibinfo {title} {Long-time tails in angular
  momentum correlations},}\ }\href@noop {} {\bibfield  {journal} {\bibinfo
  {journal} {The Journal of Chemical Physics}\ }\textbf {\bibinfo {volume}
  {103}},\ \bibinfo {pages} {1582--1587} (\bibinfo {year} {1995})}\BibitemShut
  {NoStop}%
\bibitem [{\citenamefont {van~der Hoef}\ and\ \citenamefont
  {Frenkel}(1995)}]{van_der_hoef_computer_1995}%
  \BibitemOpen
  \bibfield  {author} {\bibinfo {author} {\bibfnamefont {Martin~A.}\
  \bibnamefont {van~der Hoef}}\ and\ \bibinfo {author} {\bibfnamefont {Daan}\
  \bibnamefont {Frenkel}},\ }\bibfield  {title} {\enquote {\bibinfo {title}
  {Computer simulations of long-time tails: {What}'s new?}}\ }\href@noop {}
  {\bibfield  {journal} {\bibinfo  {journal} {Transport Theory and Statistical
  Physics}\ }\textbf {\bibinfo {volume} {24}},\ \bibinfo {pages} {1227--1247}
  (\bibinfo {year} {1995})}\BibitemShut {NoStop}%
\bibitem [{\citenamefont {Lowe}\ and\ \citenamefont
  {Frenkel}(1995)}]{lowe_super_1995}%
  \BibitemOpen
  \bibfield  {author} {\bibinfo {author} {\bibfnamefont {C.~P.}\ \bibnamefont
  {Lowe}}\ and\ \bibinfo {author} {\bibfnamefont {D.}~\bibnamefont {Frenkel}},\
  }\bibfield  {title} {\enquote {\bibinfo {title} {The super long-time decay of
  velocity fluctuations in a two-dimensional fluid},}\ }\href@noop {}
  {\bibfield  {journal} {\bibinfo  {journal} {Physica A: Statistical Mechanics
  and its Applications}\ }\textbf {\bibinfo {volume} {220}},\ \bibinfo {pages}
  {251--260} (\bibinfo {year} {1995})}\BibitemShut {NoStop}%
\bibitem [{\citenamefont {Boussinesq}(1901)}]{boussinesq_theorie_1901}%
  \BibitemOpen
  \bibfield  {author} {\bibinfo {author} {\bibfnamefont {J}~\bibnamefont
  {Boussinesq}},\ }\href@noop {} {\emph {\bibinfo {title} {Th\'eorie
  {Analytique} de la {Chaleur} ({Vol}. 2)}}}\ (\bibinfo {year}
  {1901})\BibitemShut {NoStop}%
\bibitem [{\citenamefont {Chow}\ and\ \citenamefont
  {Hermans}(1972)}]{chow_effect_1972}%
  \BibitemOpen
  \bibfield  {author} {\bibinfo {author} {\bibfnamefont {T.~S.}\ \bibnamefont
  {Chow}}\ and\ \bibinfo {author} {\bibfnamefont {J.~J.}\ \bibnamefont
  {Hermans}},\ }\bibfield  {title} {\enquote {\bibinfo {title} {Effect of
  {Inertia} on the {Brownian} {Motion} of {Rigid} {Particles} in a {Viscous}
  {Fluid}},}\ }\href@noop {} {\bibfield  {journal} {\bibinfo  {journal} {The
  Journal of Chemical Physics}\ }\textbf {\bibinfo {volume} {56}},\ \bibinfo
  {pages} {3150} (\bibinfo {year} {1972})}\BibitemShut {NoStop}%
\bibitem [{\citenamefont {Franosch}\ \emph {et~al.}(2011)\citenamefont
  {Franosch}, \citenamefont {Grimm}, \citenamefont {Belushkin}, \citenamefont
  {Mor}, \citenamefont {Foffi}, \citenamefont {Forr\'o},\ and\ \citenamefont
  {Jeney}}]{franosch_resonances_2011}%
  \BibitemOpen
  \bibfield  {author} {\bibinfo {author} {\bibfnamefont {Thomas}\ \bibnamefont
  {Franosch}}, \bibinfo {author} {\bibfnamefont {Matthias}\ \bibnamefont
  {Grimm}}, \bibinfo {author} {\bibfnamefont {Maxim}\ \bibnamefont
  {Belushkin}}, \bibinfo {author} {\bibfnamefont {Flavio~M.}\ \bibnamefont
  {Mor}}, \bibinfo {author} {\bibfnamefont {Giuseppe}\ \bibnamefont {Foffi}},
  \bibinfo {author} {\bibfnamefont {L\'aszl\'o}\ \bibnamefont {Forr\'o}}, \
  and\ \bibinfo {author} {\bibfnamefont {Sylvia}\ \bibnamefont {Jeney}},\
  }\bibfield  {title} {\enquote {\bibinfo {title} {Resonances arising from
  hydrodynamic memory in {Brownian} motion},}\ }\href@noop {} {\bibfield
  {journal} {\bibinfo  {journal} {Nature}\ }\textbf {\bibinfo {volume} {478}},\
  \bibinfo {pages} {85--88} (\bibinfo {year} {2011})}\BibitemShut {NoStop}%
\bibitem [{\citenamefont {Lesnicki}\ \emph {et~al.}(2016)\citenamefont
  {Lesnicki}, \citenamefont {Vuilleumier}, \citenamefont {Carof},\ and\
  \citenamefont {Rotenberg}}]{lesnicki_molecular_2016}%
  \BibitemOpen
  \bibfield  {author} {\bibinfo {author} {\bibfnamefont {Dominika}\
  \bibnamefont {Lesnicki}}, \bibinfo {author} {\bibfnamefont {Rodolphe}\
  \bibnamefont {Vuilleumier}}, \bibinfo {author} {\bibfnamefont {Antoine}\
  \bibnamefont {Carof}}, \ and\ \bibinfo {author} {\bibfnamefont {Benjamin}\
  \bibnamefont {Rotenberg}},\ }\bibfield  {title} {\enquote {\bibinfo {title}
  {Molecular {Hydrodynamics} from {Memory} {Kernels}},}\ }\href@noop {}
  {\bibfield  {journal} {\bibinfo  {journal} {Physical Review Letters}\
  }\textbf {\bibinfo {volume} {116}},\ \bibinfo {pages} {147804} (\bibinfo
  {year} {2016})}\BibitemShut {NoStop}%
\bibitem [{\citenamefont {Alder}\ and\ \citenamefont
  {Wainwright}(1970)}]{alder_decay_1970}%
  \BibitemOpen
  \bibfield  {author} {\bibinfo {author} {\bibfnamefont {B.~J.}\ \bibnamefont
  {Alder}}\ and\ \bibinfo {author} {\bibfnamefont {T.~E.}\ \bibnamefont
  {Wainwright}},\ }\bibfield  {title} {\enquote {\bibinfo {title} {Decay of the
  velocity autocorrelation function},}\ }\href@noop {} {\bibfield  {journal}
  {\bibinfo  {journal} {Physical review A}\ }\textbf {\bibinfo {volume} {1}},\
  \bibinfo {pages} {18} (\bibinfo {year} {1970})}\BibitemShut {NoStop}%
\bibitem [{\citenamefont {Hasimoto}(1959)}]{hasimoto_periodic_1959}%
  \BibitemOpen
  \bibfield  {author} {\bibinfo {author} {\bibfnamefont {H.}~\bibnamefont
  {Hasimoto}},\ }\bibfield  {title} {\enquote {\bibinfo {title} {On the
  periodic fundamental solutions of the {Stokes} equations and their
  application to viscous flow past a cubic array of spheres},}\ }\href@noop {}
  {\bibfield  {journal} {\bibinfo  {journal} {Journal of Fluid Mechanics}\
  }\textbf {\bibinfo {volume} {5}},\ \bibinfo {pages} {317--328} (\bibinfo
  {year} {1959})}\BibitemShut {NoStop}%
\bibitem [{\citenamefont {Succi}(2001)}]{SucciBook}%
  \BibitemOpen
  \bibfield  {author} {\bibinfo {author} {\bibfnamefont {S.}~\bibnamefont
  {Succi}},\ }\href@noop {} {\emph {\bibinfo {title} {The Lattice Boltzmann
  Equation for Fluid Dynamics and Beyond}}}\ (\bibinfo  {publisher} {Oxford
  University Press},\ \bibinfo {year} {2001})\BibitemShut {NoStop}%
\bibitem [{Sup()}]{SuppMat}%
  \BibitemOpen
  \href@noop {} {\enquote {\bibinfo {title} {See supplemental material at [url
  will be inserted by publisher] for the list of simulated systems.}}\
  }\BibitemShut {NoStop}%
\bibitem [{\citenamefont {Dellar}(2001)}]{dellar_bulk_2001}%
  \BibitemOpen
  \bibfield  {author} {\bibinfo {author} {\bibfnamefont {Paul~J.}\ \bibnamefont
  {Dellar}},\ }\bibfield  {title} {\enquote {\bibinfo {title} {Bulk and shear
  viscosities in lattice {Boltzmann} equations},}\ }\href@noop {} {\bibfield
  {journal} {\bibinfo  {journal} {Physical Review E}\ }\textbf {\bibinfo
  {volume} {64}} (\bibinfo {year} {2001})}\BibitemShut {NoStop}%
\bibitem [{\citenamefont {Li}\ and\ \citenamefont
  {Shan}(2011)}]{li_lattice_2011}%
  \BibitemOpen
  \bibfield  {author} {\bibinfo {author} {\bibfnamefont {Y.}~\bibnamefont
  {Li}}\ and\ \bibinfo {author} {\bibfnamefont {X.}~\bibnamefont {Shan}},\
  }\bibfield  {title} {\enquote {\bibinfo {title} {Lattice {Boltzmann} method
  for adiabatic acoustics},}\ }\href@noop {} {\bibfield  {journal} {\bibinfo
  {journal} {Philosophical Transactions of the Royal Society A: Mathematical,
  Physical and Engineering Sciences}\ }\textbf {\bibinfo {volume} {369}},\
  \bibinfo {pages} {2371--2380} (\bibinfo {year} {2011})}\BibitemShut {NoStop}%
\bibitem [{\citenamefont {Viggen}(2011)}]{viggen_viscously_2011}%
  \BibitemOpen
  \bibfield  {author} {\bibinfo {author} {\bibfnamefont {Erlend~Magnus}\
  \bibnamefont {Viggen}},\ }\bibfield  {title} {\enquote {\bibinfo {title}
  {Viscously damped acoustic waves with the lattice {Boltzmann} method},}\
  }\href@noop {} {\bibfield  {journal} {\bibinfo  {journal} {Philosophical
  Transactions of the Royal Society of London A: Mathematical, Physical and
  Engineering Sciences}\ }\textbf {\bibinfo {volume} {369}},\ \bibinfo {pages}
  {2246--2254} (\bibinfo {year} {2011})}\BibitemShut {NoStop}%
\bibitem [{\citenamefont {K\"uhne}\ \emph {et~al.}(2009)\citenamefont
  {K\"uhne}, \citenamefont {Krack},\ and\ \citenamefont
  {Parrinello}}]{kuhne_static_2009}%
  \BibitemOpen
  \bibfield  {author} {\bibinfo {author} {\bibfnamefont {Thomas~D.}\
  \bibnamefont {K\"uhne}}, \bibinfo {author} {\bibfnamefont {Matthias}\
  \bibnamefont {Krack}}, \ and\ \bibinfo {author} {\bibfnamefont {Michele}\
  \bibnamefont {Parrinello}},\ }\bibfield  {title} {\enquote {\bibinfo {title}
  {Static and {Dynamical} {Properties} of {Liquid} {Water} from {First}
  {Principles} by a {Novel} {Car}-{Parrinello}-like {Approach}},}\ }\href@noop
  {} {\bibfield  {journal} {\bibinfo  {journal} {Journal of Chemical Theory and
  Computation}\ }\textbf {\bibinfo {volume} {5}},\ \bibinfo {pages} {235--241}
  (\bibinfo {year} {2009})}\BibitemShut {NoStop}%
\bibitem [{\citenamefont {Hagen}\ \emph {et~al.}(1997)\citenamefont {Hagen},
  \citenamefont {Pagonabarraga}, \citenamefont {Lowe},\ and\ \citenamefont
  {Frenkel}}]{hagen_algebraic_1997}%
  \BibitemOpen
  \bibfield  {author} {\bibinfo {author} {\bibfnamefont {M.~H.~J.}\
  \bibnamefont {Hagen}}, \bibinfo {author} {\bibfnamefont {I.}~\bibnamefont
  {Pagonabarraga}}, \bibinfo {author} {\bibfnamefont {C.~P.}\ \bibnamefont
  {Lowe}}, \ and\ \bibinfo {author} {\bibfnamefont {Daan}\ \bibnamefont
  {Frenkel}},\ }\bibfield  {title} {\enquote {\bibinfo {title} {Algebraic decay
  of velocity fluctuations in a confined fluid},}\ }\href@noop {} {\bibfield
  {journal} {\bibinfo  {journal} {Physical review letters}\ }\textbf {\bibinfo
  {volume} {78}},\ \bibinfo {pages} {3785} (\bibinfo {year}
  {1997})}\BibitemShut {NoStop}%
\bibitem [{\citenamefont {Huang}\ and\ \citenamefont
  {Szlufarska}(2015)}]{huang_effect_2015}%
  \BibitemOpen
  \bibfield  {author} {\bibinfo {author} {\bibfnamefont {Kai}\ \bibnamefont
  {Huang}}\ and\ \bibinfo {author} {\bibfnamefont {Izabela}\ \bibnamefont
  {Szlufarska}},\ }\bibfield  {title} {\enquote {\bibinfo {title} {Effect of
  interfaces on the nearby {Brownian} motion},}\ }\href@noop {} {\bibfield
  {journal} {\bibinfo  {journal} {Nature Communications}\ }\textbf {\bibinfo
  {volume} {6}},\ \bibinfo {pages} {8558} (\bibinfo {year} {2015})}\BibitemShut
  {NoStop}%
\end{thebibliography}
%

\begin{acknowledgments}
The authors are grateful to Jean-Pierre Hansen and Lyd\'eric Bocquet for useful
discussions. AJA and BR acknowledge financial support from the French Agence Nationale de
la Recherche (ANR) under grant ANR-15-CE09-0013-01.
\end{acknowledgments}

\end{document}